\documentclass[fleqn,10pt]{wlscirep}
\usepackage[utf8]{inputenc}
\usepackage[T1]{fontenc}

\usepackage{subcaption}

\usepackage{scalerel}
\def\up{{\scalerel*{\uparrow}{1}}}
\def\dn{{\scalerel*{\downarrow}{1}}}
\def\ud{{\scalerel*{\updownarrow}{1}}}

\usepackage{soul}

\usepackage{bm}
\usepackage{array}
\usepackage{physics}
\usepackage{mathtools,mathrsfs,amsfonts}
\newcommand{\bb}[1]{\pmb{#1}}
\usepackage{textgreek}
\renewcommand{\Pi}{\text{\textPi}}
\DeclareRobustCommand{\chi}{{\mathpalette\irchi\relax}}
\newcommand{\irchi}[2]{\raisebox{\depth}{$#1\text{\textchi}$}}

\newcommand{\detlef}[1]{\textcolor{black} {#1}}
\newcommand{\mike}{\textcolor{black}}
\newcommand{\SD}{\textcolor{black}}

\title{Arrival Time Distributions of Spin-1/2 Particles}

\author{Siddhant Das~~~and~~~Detlef D\"{u}rr}
\affil{Mathematisches Institut, Ludwig-Maximilians-Universitat M\"{u}nchen, Theresienstr. 39, D-80333 M\"{u}nchen, Germany}

\affil{Siddhant.Das@physik.uni-muenchen.de\\duerr@mathematik.uni-muenchen.de}

\begin{abstract}
The arrival time statistics of spin-1/2 particles governed by Pauli's equation, and defined by their Bohmian trajectories, show unexpected and very well articulated features. Comparison with other proposed statistics of arrival times that arise from either the usual (convective) quantum flux or from semiclassical considerations suggest testing the notable deviations in an arrival time experiment, thereby probing the predictive power of Bohmian trajectories. The suggested experiment, including the preparation of the wave functions, could be done with present-day experimental technology. 
\end{abstract}

\begin{document}

\flushbottom
\maketitle

\thispagestyle{empty}

\section*{Introduction}

In non-relativistic quantum mechanics, the probability of finding a particle in a small spatial volume $\mathrm{d}^3r$ around position $\bb{r}$ at a fixed time $t$ is given by Born's rule $|\Psi(\bb{r},t)|^2\mathrm{d}^3r$, where $\Psi(\bb{r},t)$ is the wave function of the particle. This formula is experimentally well established. However, a formula for the probability of finding the particle at a fixed point $\bb{r}$ between times $t$ and $t+\mathrm{d}t$ is the matter of an ongoing debate \cite{MUGA,MUGA1,Allcock1,AhBohm,Vona,DDGZ}. Let us consider a typical time of arrival experiment, in which a particle is initially trapped in a region $\Sigma\subset\mathbb{R}^3$, e.g., the interior of a potential well. The trap is released at, say, $t=0$, allowing the particle to propagate freely in space. If the trapping potential is deep enough, the wave function of the particle at this instant, $\Psi_0(\bb{r})\equiv\Psi(\bb{r},t=0)$, practically vanishes outside the region $\Sigma$. Particle detectors placed on the boundary $\partial\Sigma$ measure the time of arrival of the particle, denoted by $\tau$. If the experiment is repeated many times, the recorded arrival times are random, even if the initial wave function of the particle ($\Psi_0$) is kept unchanged in each experiment. What is the probability distribution of arrival times $\Pi^{\Psi_0}(\tau)$ as a functional depending on $\Psi_0$ and on $\partial\Sigma$?

Measurement in quantum mechanics \mike{has become} in recent decades a tricky notion. \mike{Traditionally,} measurement outcomes were solely associated with observables, represented by self-adjoint operators on the Hilbert space of the measured system. \mike{However,} it has long been known that for \mike{time measurements}, \mike{such as} arrival times, no such observable exists \SD{\cite[\S\,8.5]{MUGA1}}. In fact, the notion of self-adjoint operators defining quantum observables does not apply to many other experiments as well. \mike{To remedy this situation,} the notion of \SD{an} observable was generalized to positive operator valued measures (POVMs), and there are various suggestions for arrival time POVMs (see \cite{Vona1,Rodi} for a discussion). In Dürr et al.,\cite{DGZOperators} all measurements \mike{describable} by POVMs were called linear measurements. \SD{Another} class of measurements have also been performed\SD{, the so-called} weak measurements,\cite{Kocsis} \SD{which are nonlinear in the sense of\cite{DGZOperators}}. So far, however, no \SD{theoretical} predictions for the arrival time distribution $\Pi^{\Psi_0}(\tau)$ \SD{have been backed up by experiments} (see \cite{MUGA,MUGA1,Yearsley,Allcock2,Allcock3,Rodi} for various \SD{proposals}). \SD{On the other hand, recent} `attoclock' experiments (\SD{claimed to be} measuring the tunnel delay time of ionized electrons) have shown some of the theoretical \SD{ideas} to be empirically inadequate \cite{Keller,Lundsmann}.

One problem with arrival time measurements is that detection events are based on interactions of the detectors with the detected particle, which may disturb its wave function in an uncontrollable way, leading to backscattering and in extreme cases to the quantum Zeno effect. While this is a valid concern, we note that the double-slit experiment (mentioned also below) is an example where the distribution of arrival positions of the particle on the detector screen (the ubiquitous interference picture) is analyzed without any reference to the presence of the detector. Note well that the particles strike the detector surface at \emph{random times}, a fact blissfully ignored in the usual discussions of the double-slit experiment--and there are good reasons why that is justified. We expect that in the experiment proposed in this paper the same will be true, i.e., the detection event should not be drastically disturbed by the presence of the detector. Our expectation is based on our results, namely, on the very striking articulated features of the computed arrival time distributions, which should survive mild disturbances. 


A further problem is that the notion of arrival time is most naturally connected with that of particle trajectories, an idea which is hard to concretize in the orthodox interpretation of quantum mechanics. Bohmian mechanics (or de Broglie-Bohm pilot wave theory) is a quantum theory \mike{(and not simply an alternative interpretation of quantum mechanics)} where particles move on well defined smooth trajectories, hence it is naturally suited for computing arrival times of a particle. See Kocsis et al. \cite{Kocsis} for a {\textit weak} measurement of average quantum trajectories, which can indeed be seen as Bohmian trajectories. Bohmian mechanics \mike{has been} proven to be \textit{empirically equivalent to standard quantum mechanics, wherever the latter is unambiguous} \detlef{(e.g., in position \SD{and momentum} measurements)}\cite{BohmHiley,DGZOperators}. It has been shown that Bohmian mechanics provides \SD{in (far field) scattering situations} \emph{ideal} arrival time statistics for spin-0 particles\cite{DDGZ}, via the quantum flux (\SD{or the probability current}) $\bb{\mathrm{J}}$:
\begin{equation}\label{qufl}
\Pi^{\Psi_0}(\tau)=\int_{\partial\Sigma}\bb{\mathrm{J}}(\bb{r},\tau)\!\cdot\!\mathrm{d}\bb{\text{s}}\qquad(\eqqcolon\Pi_{\texttt{qf}}(\tau)).
\end{equation}
Only in scattering situations (i.e., when the detector surface $\partial\Sigma$ is far away from the support of the initial wave function $\Psi_0$), is the surface integral in \eqref{qufl} demonstrably positive\cite{DDGZ}, otherwise it cannot be interpreted as a probability density (see\cite{Vona1} for a discussion of POVMs versus flux statistics). Although often not recognized or emphasized in textbooks, it is the quantum flux $\bb{\mathrm{J}}$, integrated over time, that yields the double-slit interference pattern of arrival {\emph {positions}} of particles on the screen--–as there is no \emph{given} time at which the particles arrive at the screen.

Here, we must mention as well another flux based (Bohmian) arrival time distribution derived by C. R. Leavens (Eq. (9) of\cite{Leavens}), which is valid \emph{only} in one space dimension (as discussed in \cite[Ch. 5]{MUGA}). Therefore, it is not applicable for particles with spin-1/2, except perhaps in some idealized situations. In this paper, we propose a Bohmian formula for the distribution \mike{of \emph{first} arrival times} \SD{of a spin-1/2 particle (Eq. \eqref{BohmTime} below)}, which \SD{may} be referred to as \mike{an} \textit{ideal} or \textit{intrinsic} distribution, since it is \SD{formulated without referring to} any \SD{particular} measurement device, just as equation \eqref{qufl}. In fact, the spin-1/2 analogue of \eqref{qufl} becomes a special case of \eqref{BohmTime}, whenever the so-called \emph{current positivity} condition\detlef{\cite{DDGZ}} is met. \SD{Evaluating our formula numerically} for a specific, \mike{carefully} chosen experiment, we find that the resulting arrival time distributions \detlef{show drastic and unexpected changes when control parameters are \SD{varied}.} Since the predicted distributions show such interesting and significant behavior, we suggest that \SD{the proposed} experiment be performed to test the predictive power of Bohmian mechanics for spin-1/2 particles. 

\detlef{One may legitimately ask, how can the idea of an ideal first arrival time distribution be entertained at all?} Our standpoint on this is as follows: Given the \mike{relative simplicity of making the Bohmian prediction for the ideal \SD{arrival time} distribution}, and given the ambiguity of quantum mechanical proposals mentioned above, why not \mike{just} do the experiment to check \mike{it}? \mike{However} the experimental results turn out, they provide in any case valuable experimental data that would enrich our understanding of quantum mechanics and of Bohmian mechanics as well. \detlef{A word of clarification is, however, in order: we do not construe \mike{in our work any \textit{contradiction}} to quantum mechanics, since an unambiguous answer to the question of arrival times has not been given within this framework. The main advantage of Bohmian mechanics is the clear picture of reality, independent of observation, it provides and which in the problem of arrival times allows an unambiguous answer in contrast to the answers given in quantum mechanics, \SD{so far}.}

\begin{figure}[!ht]
\centering
\includegraphics[scale=1.3]{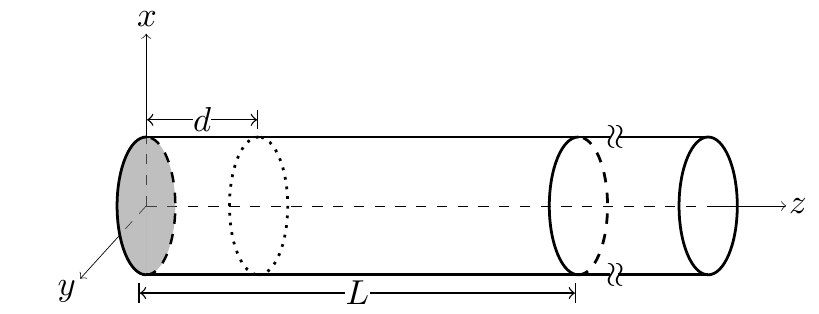}
\caption{Schematic drawing of the experimental setup. The barrier at $d$ is switched off at $t=0$ and arrival times are monitored at $z=L$.}\label{waveguide}
\end{figure}

We \mike{proceed now to a description of the proposed experiment.} A spin-1/2 particle of mass $m$ is constrained to move within a semi-infinite cylindrical waveguide (Fig. \ref{waveguide}). Initially, it is trapped between the end face of the waveguide and an impenetrable potential barrier placed at a distance $d$. At the start of the experiment, the particle is prepared in a ground state $\Psi_0$ of this cylindrical box, then the barrier at $d$ is suddenly switched off, allowing the particle to propagate freely within the waveguide. The arrival surface $\partial\Sigma$ is the \emph{plane} situated at distance $L\,\,(>d)$ from the end face of the waveguide. We then compute  \textit{numerically} from the Bohmian equations of motion  how long it takes for the particle to arrive at $\partial\Sigma$, and determine the empirical distribution $\Pi^{\Psi_0}_{\texttt{Bohm}}(\tau)$ from \textit{typical} trajectories (trajectories whose initial points are randomly drawn from the Born $|\Psi_0|^2$ distribution) for different initial ground state wave functions.
\par The Bohmian equations for spin-1/2 particles are as follows:
The wave function $\Psi(\bb{r},t)$ is a two-component complex-valued spinor solution of the Pauli equation
\begin{align}\label{pauli}
   i\hbar\frac{\partial}{\partial t}&\Psi(\bb{r},t)=-\frac{\hbar^2}{2m}\left(\bb{\sigma}\!\cdot\!\bb{\nabla}\right)^2\Psi(\bb{r},t)+V(\bb{r},t)\Psi(\bb{r},t),
\end{align}
with given initial condition $\Psi_0(\bb{r})$. Here, $V(\bb{r},t)$ is an external potential, and $\bb{\sigma}=\sigma_x\,\hat{\bb{x}}+\sigma_y\,\hat{\bb{y}}+\sigma_z\,\hat{\bb{z}}$ is a 3-vector of Pauli spin matrices. The quantum continuity equation for the Pauli equation reads \cite{J1,J2}
\begin{align}\label{quconPauli}
    \frac{\partial |\Psi|^2}{\partial t}=\frac{\hbar}{m}\,\bb{\nabla}\!\cdot\!\Big(\mathrm{Im}[\Psi^{\dagger}\bb{\nabla}\Psi]+\frac{1}{2}\,\bb{\nabla}\!\times\!(\Psi^{\dagger}\bb{\sigma}\Psi)\Big)=\bb{\nabla}\!\cdot\!\bb{\mathrm{J}}_{\texttt{Pauli}}\eqqcolon\bb{\nabla}\!\cdot\!\left(\bb v^\Psi_{\texttt{Bohm}} |\Psi|^2\right),
\end{align}
where $\Psi^{\dagger}$ is the adjoint of $\Psi$, and $|\Psi|^2=\Psi^{\dagger}\Psi$. The rightmost equality of \eqref{quconPauli} defines the Bohmian velocity field $\bb{v}^\Psi_{\texttt{Bohm}}=\bb{\mathrm{J}}_{\texttt{Pauli}}/ |\Psi|^2$. The term $\frac{\hbar}{m}\mathrm{Im}[\Psi^{\dagger}\bb{\nabla}\Psi]$ is the so-called \emph{convective} flux, while $\frac{\hbar}{2m}\,\bb{\nabla}\!\times\!(\Psi^{\dagger}\bb{\sigma}\Psi)$ is the \emph{spin} flux. The alert reader will recognize that the spin flux is divergenceless, hence one may argue that neither the flux $\bb{\mathrm{J}}_{\texttt{Pauli}}$ nor the Bohmian velocity are uniquely defined. However, observing that the Pauli equation and its flux emerge as non-relativistic limits of the Dirac equation and the Dirac flux, respectively, which are \emph{unique} \cite{Holland2003,Holland,HollandPhilippidis}, one is led directly to the current and Bohmian velocity given here.

The Bohmian trajectories are integral curves of the velocity field $\bb v^\Psi_{\texttt{Bohm}}$, hence the Bohmian \textit{guidance law} reads
\begin{align}\label{guidance}
\frac{\mathrm{d}}{\mathrm{d}t}\bb{R}(t)=\bb v^\Psi_{\texttt{Bohm}}(\bb{R}(t),t)=\frac{\hbar}{m}\,\mathrm{Im}\!\left[\frac{\Psi^{\dagger}\bb{\nabla}\Psi}{\Psi^{\dagger}\Psi}\right]\!\left(\bb{R}(t),t\right)+\frac{\hbar}{2m}\left[\frac{\bb{\nabla}\!\times\!\left(\Psi^{\dagger}\bb{\sigma}\Psi\right)}{\Psi^{\dagger}\Psi}\right]\!\left(\bb{R}(t),t\right).
\end{align}
\noindent Here, $\bb{R}(t)$ is the position of the particle at time $t$. In Bohm's theory, the spin-1/2 particle has no degrees of freedom other than those specifying its position in space, so spin is not an \emph{extra} degree of freedom. Thus, all quantum mechanical phenomena attributed to spin (such as the deflection of particles in the Stern-Gerlach experiment) arise solely from the non-linear equation of motion \cite{HDK1,HDK2}. 
The guidance law \eqref{guidance} is time reversal invariant and its right-hand side transforms as a velocity under Galilean transformations (see \cite{Holland,BohmHiley} for further discussion). We integrate Eq. \eqref{guidance} for a statistical ensemble of $|\Psi_0|^2$-distributed initial particle positions $\bb{R}(0)$ (see \cite{DGZBorn} for a justification).
\section*{Methods}
Our computations and results are for the following setup: Let the cylindrical waveguide be mounted on the $xy-$plane of a right-handed orthogonal coordinate system, the axis of the cylinder defining the $z-$axis (Fig.\ref{waveguide}). Employing cylindrical coordinates $\bb{r}\equiv(\rho,\phi,z)$, we model the potential field of the waveguide as $V(\bb{r},t)=V_{\perp}(\rho)+V_{\parallel}(z,t)$, where $V_{\perp}(\rho)=\frac{1}{2}m\,\omega^2\rho^2$ is a transverse confining potential, and $V_{\parallel}(z,t)=v(z)+\theta(-t)v(d-z)$ is a time dependent axial potential comprised of an impenetrable hard-wall at $z=0$, viz.,
\begin{equation}
    v(z)=\begin{cases}
       \infty~&z\leq0\\
       0~&z>0
\end{cases},
\end{equation}
and another impenetrable potential barrier $v(d-z)$ that is switched off at $t=0$. The wall at $z=0$ is the end face of the waveguide and $\theta(x)$ is Heaviside's step function. Fortunately, near perfect harmonic confinements can be realized in conventional (ultrahigh vacuum) Penning traps, which can trap single electrons \cite{Dehmelt,Joseph} and protons \cite{Ulmer} over a wide range of trapping frequencies. For electrons, typical waveguide parameters read: $L\approx6-10\,\texttt{mm}$, $\omega\approx10^9-10^{11}\,\texttt{rad/s}$ \cite{Dehmelt, Dehmelt1}. In \cite{DMD} we give a detailed analysis of the proposed experiment for a quadrupole ion trap (Paul trap) waveguide.
\par The particle is prepared in a ground state of the cylindrical box at $t=0$, which can be written as $\Psi_0(\bb{r})=\psi_0(\bb{r})\chi$, where (setting $\hbar=m=d=1$),
\begin{align}\label{space}
    \psi_0(\bb{r})&=\sqrt{\frac{2\omega}{\!\pi}}\,\theta(z)\theta(1-z)\sin(\pi z)\exp\!\left(-\frac{\omega}{2}\rho^2\right)
\end{align}
is the spatial part of the wave function, and
\begin{equation}\label{spin}
    \chi=\left(\!\!\begin{array}{c}\cos(\alpha/2)\\\sin(\alpha/2)\,e^{i\beta}\\\end{array}\!\!\right),\quad0\leq\alpha\leq\pi,\quad0\leq\beta<2\pi,
\end{equation}
is a normalized Bloch spinor ($\chi^{\dagger}\chi=1$). Fixing $\alpha$ and $\beta$, we obtain different ground state wave functions. For instance, $\alpha=0~(\pi)$ gives the spin-up (spin-down) ground state wave function, usually denoted by $\Psi_\up~(\Psi_\dn)$, while $\alpha=\frac{\pi}{2}$ and $\beta=0$ yields the so-called up-down ground state wave function $\Psi_\ud=\frac{1}{\sqrt{2}}(\Psi_\up+\Psi_\dn)$. We refer to $\alpha$ and $\beta$ as \textit{spin orientation angles}, because they specify the orientation of the ``spin vector'' $\bb{\text{s}}\coloneqq\frac{1}{2}(\Psi_0^{\dagger}\,\bb{\sigma}\Psi_0^{\textcolor{white}{\dagger}})/|\Psi_0|^2$, given by
\begin{align*}
\bb{\text{s}}=\frac{1}{2}\big(\!\sin\alpha\,\cos\beta\,\hat{\bb{x}}+\sin\alpha\,\sin\beta\,\hat{\bb{y}}+\cos\alpha\,\hat{\bb{z}}\big).
\end{align*}
\par The instant the barrier is switched off, the wave function spreads dispersively, filling the volume of the waveguide. The particle moves according to \eqref{guidance}  on  the Bohmian trajectory $\bb{R}(t)=R(t)\,[\,\cos{\Phi(t)}\,\hat{\bb{x}} + \sin{\Phi(t)}\,\hat{\bb{y}}\,] + Z(t)\,\hat{\bb{z}}$.  In this choice of coordinates the first arrival time of a trajectory starting at $\bb{R}(0)$  and arriving at $z=L$ is  
\begin{equation}\label{defn}
    \tau(\bb{R}(0))=\mathrm{inf}\!\left\{t~|~Z(t,\bb{R}(0))=L,~\bb{R}(0)\in\mathrm{supp}(\Psi_0)\right\},
\end{equation}
where $Z(t,\bb{R}(0))\equiv Z(t)$ is the $z-$coordinate of the particle at time $t$, and $\mathrm{supp}(\Psi_0)$ denotes the support of the initial wave function (the interior of the cylindrical box). Since the initial position $\bb{R}(0)$ is $|\Psi_0|^2$-distributed\cite{DGZBorn}, the distribution of $\tau(\bb{R}(0))$ is given by
\begin{equation}\label{BohmTime}
    \Pi_{\texttt{Bohm}}^{\Psi_0}(\tau)=\int_{\mathrm{supp}(\Psi_0)}\kern-1em\mathrm{d}^3\bb{R}(0)~\delta\big(\tau(\bb{R}(0))-\tau\big)\,|\Psi_0|^2(\bb{R}(0))\,.\vspace{2pt}
\end{equation}
In general, there is no closed form expression for \eqref{defn}, hence the integral in \eqref{BohmTime} cannot be evaluated analytically. However, if the Bohmian trajectories cross $\partial\Sigma$ \textit{at most} once, or in other words if the quantum flux $\bb{\mathrm{J}}_{\texttt{Pauli}}$ is outward directed at \textit{every point} of $\partial\Sigma$, at \textit{all times} (also referred to as the {\textit {current positivity}} condition), then $\Pi_{\texttt{Bohm}}^{\Psi_0}(\tau)$ reduces to the integrated quantum flux $\Pi_\texttt{qf}(\tau)$, Eq. \eqref{qufl} with the Pauli current $\bb{\mathrm{J}}_{\texttt{Pauli}}$ replacing $\bb{\mathrm{J}}$.\cite{Vona,DDGZ}

Note well that by the very meaning of the quantum flux, \eqref{qufl} is a natural guess for the arrival time distribution from the point of view of standard quantum mechanics as well \cite{Vona}. However, \eqref{qufl} makes sense \emph{only} if the left-hand side is positive, which need not be the case. Of course, if the current positivity condition holds, the left-hand side of \eqref{qufl} $\geq0$, and $\Pi_{\texttt{qf}}$ becomes a special case of \eqref{BohmTime}. Generally, this condition does \textit{not} hold, in which case one computes $\Pi_{\texttt{Bohm}}^{\Psi_0}(\tau)$ \textit{numerically} from a large number of Bohmian trajectories.
\section*{Results and Discussion}
(i) For the spin-up ($\alpha=0$) and spin-down ($\alpha=\pi$) wave functions the arrival time distribution $\Pi^{\Psi_0}_{\texttt{Bohm}}(\tau)$ coincides with the quantum flux expression \eqref{qufl}, since in these cases the current positivity condition is satisfied. Moreover, in these cases $\bb{\mathrm{J}}_{\texttt{Pauli}}(\bb{r},\tau)$ can be replaced by the convective flux $\frac{\hbar}{m}\mathrm{Im}[\Psi^{\dagger}\bb{\nabla}\Psi]$ in \eqref{qufl}. The resulting distribution has a heavy tail $\sim\tau^{-4}$ as $\tau\to\infty$. (ii) For other initial wave functions $\Pi_{\texttt{Bohm}}^{\Psi_0}(\tau)$ differs from \eqref{qufl} and falls off faster than $\sim\tau^{-4}$. For any initial ground state wave function, the arrival time distribution displays an infinite sequence of \textit{self-similar} lobes below $\tau=\frac{mdL}{2\pi\hbar}$ (see Fig. \ref{w1000} below), which diminish in size as $\tau\to 0$. These lobes mirror typical wave function evolution when suddenly released to spread freely into the volume of the waveguide (see also \cite{Moshinsky}). (iii) If the initial wave function is an \textit{equal superposition} of the spin-up and spin-down wave functions ($\alpha=\frac{\pi}{2}$), the arrival time distribution pinches off at a maximum arrival time $\tau_{\texttt{max}}$, i.e., \textit{no} particle arrivals occur for $\tau>\tau_{\texttt{max}}$. Moreover an even more striking manifestation of the lobes can be seen: characteristic ``no-arrival \textit{windows}'' appear between the smaller lobes, inside which the arrival time distribution is \textit{zero}. (iv) Time of flight measurements refer in general to semiclassical expressions based on the momentum distribution. Our distributions deviate significantly from this alleged semiclassical formula.
\par A few details concerning the computation of our results are in order. First note that the Pauli equation \eqref{pauli} with initial condition $\Psi_0(\bb{r})$ can be solved in closed form, facilitating very fast numerical computation of Bohmian trajectories. The time dependent wave function takes the form $\Psi(\bb{r},t)=\psi(\bb{r},t)\chi$, where the spin part is given by \eqref{spin}, while
\begin{equation}\label{TDWF}
    \psi(\bb{r},t)=\sqrt{\frac{2\omega}{\!\pi}}\exp\!\left(-\frac{\omega}{2}\rho^2-i\omega t\right)W(z,t),
\end{equation}
where
\begin{align}\label{TEI}
    W(z,t)=\theta(z)\big[\,\mathscr{D}(z-1,t)+\mathscr{D}(1-z,t)-\mathscr{D}(1+z,t)-\mathscr{D}(-1-z,t)\,\big],
\end{align}
which we call the `time evolution integral'. In \eqref{TEI}:
\begin{align}
    \mathscr{D}(x,t)&\coloneqq\frac{e^{-i\frac{\pi^2}{2}t}}{8i}\left\{e^{i\pi x}\,\mathrm{erfc}\!\left[\frac{i^{3/2}}{\sqrt{2}}\left(\frac{x}{\sqrt{t}}-\pi\sqrt{t}\right)\right]-e^{-i\pi x}\,\mathrm{erfc}\!\left[\frac{i^{3/2}}{\sqrt{2}}\left(\frac{x}{\sqrt{t}}+\pi\sqrt{t}\right)\right]\right\},
\end{align}
where $\mathrm{erfc}(x)$ is the complementary error function. A detailed derivation of this result will be given elsewhere \cite{prep}. 
Substituting our solution for the time dependent wave function in the guidance law \eqref{guidance}, we obtain coupled non-linear equations of motion for the spin-1/2 particle:
\begin{subequations}\label{EOM}
\begin{align}
     \dot{R}(t)&=\sin\alpha\,\sin(\Phi(t)-\beta)\,\mathrm{Re}\!\left[\frac{W'}{W}\right]\!(Z(t),t),\label{EOMr}\\
     \dot{\Phi}(t)&=\frac{\sin\alpha}{R(t)}\,\cos(\Phi(t)-\beta)\,\mathrm{Re}\!\left[\frac{W'}{W}\right]\!(Z(t),t)+\omega\cos\alpha,\label{EOMp}\\
     \dot{Z}(t)&=\mathrm{Im}\!\left[\frac{W'}{W}\right]\!(Z(t),t)+\omega\sin\alpha\,\sin(\Phi(t)-\beta)R(t),\label{EOMz}
\end{align}
\end{subequations}
where $W'=\partial W/\partial z$.
The parameters in the initial wave function are, in view of \eqref{spin}, $\alpha$ and $\beta$, so we denote $\Pi^{\Psi_0}_{\texttt{Bohm}}(\tau)\equiv\Pi^{\alpha|\beta}_{\texttt{Bohm}}(\tau)$. In fact, it is enough to consider $\beta=0$ only since 
\begin{subequations}\label{nice}
\begin{align}
    \Pi^{\alpha|\beta}_{\texttt{Bohm}}(\tau)&=\Pi^{\alpha|0}_{\texttt{Bohm}}(\tau),\label{propA}\\
    \Pi^{\alpha|\beta}_{\texttt{Bohm}}(\tau)&=\Pi^{\pi-\alpha|\beta}_{\texttt{Bohm}}(\tau).\label{propB}
\end{align}
\end{subequations}
These results are proven in \cite{prep}. We sample $N\approx 10^5$ initial positions from the $|\Psi_0|^2$ distribution, solve Eq. \eqref{EOM} numerically for each point in this ensemble, continuing until the trajectory hits $z=L$, then record the arrival time and plot the histogram for  $\Pi^{\alpha|0}_{\texttt{Bohm}}(\tau)$.
\par For the spin-up and spin-down wave functions, Eq. \eqref{EOMz} reduces to $\dot{Z}\!=\!\mathrm{Im}[W'/W](Z(t),t)$ and numerically it turns out that $\mathrm{Im}[W'/W](L,t)>0$. Hence the spin-up and spin-down trajectories cross $\partial\Sigma$ \textit{at most} once, and the first arrival time distribution (or simply the arrival time distribution) in these cases equals (cf Eq. \eqref{qufl})
\begin{equation}\label{flux}
    \Pi_{\texttt{qf}}(\tau)=2\,\mathrm{Im}[W^*(L,\tau)W'(L,\tau)].
\end{equation}
\noindent As noted, \eqref{flux} is non-negative for all values of $\tau$, and features prominently a large main lobe for $\tau>\frac{L}{2\pi}$. The main lobe falls off as
$ \left(\frac{L}{\pi}\right)^3\!\frac{4}{\tau^4}, \text{ as }\,\tau\to\infty.$ An infinite train of smaller lobes permeates the interval $0<\tau<\frac{L}{2\pi}$, which are well approximated by the formula $\frac{4\pi}{L}\,\mathrm{sinc}^2\!\left(\frac{L}{\tau}\right)$, whenever $\tau\!\ll\!L$. Apart from that we find that $\Pi_{\texttt{qf}}(\tau)$ is a function \textit{only} of the arrival distance $L$. It is also independent of the trapping frequency $\omega$, which is rather surprising. In fact, the spin-up (down) arrival time distributions are independent of the \textit{exact shape} of the transverse confining potential $V_{\perp}(\rho)$ of the waveguide as well \cite{prep}. Figure \ref{w1000} depicts our results for $L=100$ ($\approx5\,\texttt{mm}$ for a $d=50\,\text{\textmu}\texttt{m}$ trap) and $\omega=10^3$ ($\approx46.3\times10^6\,\texttt{rad/s}$). Note: We have expressed $L$, $\omega$ and $\tau$ in units of $d$, $\frac{\hbar}{md^2}$ and $\frac{md^2}{\hbar}$, respectively. For a $d=50\,\text{\textmu}\texttt{m}$ trap, known electron mass $m\approx9.11\times10^{-31}\,\texttt{kg}$ and reduced Planck's constant $\hbar\approx1.05\times10^{-34}\,\texttt{Js}$, the frequency and time units are $\approx46.3\times10^3\,\texttt{rad/s}$ and $\approx21.7\,\text{\textmu}\texttt{s}$, respectively. 

\begin{figure}[!ht]
\centering
\includegraphics[scale=1]{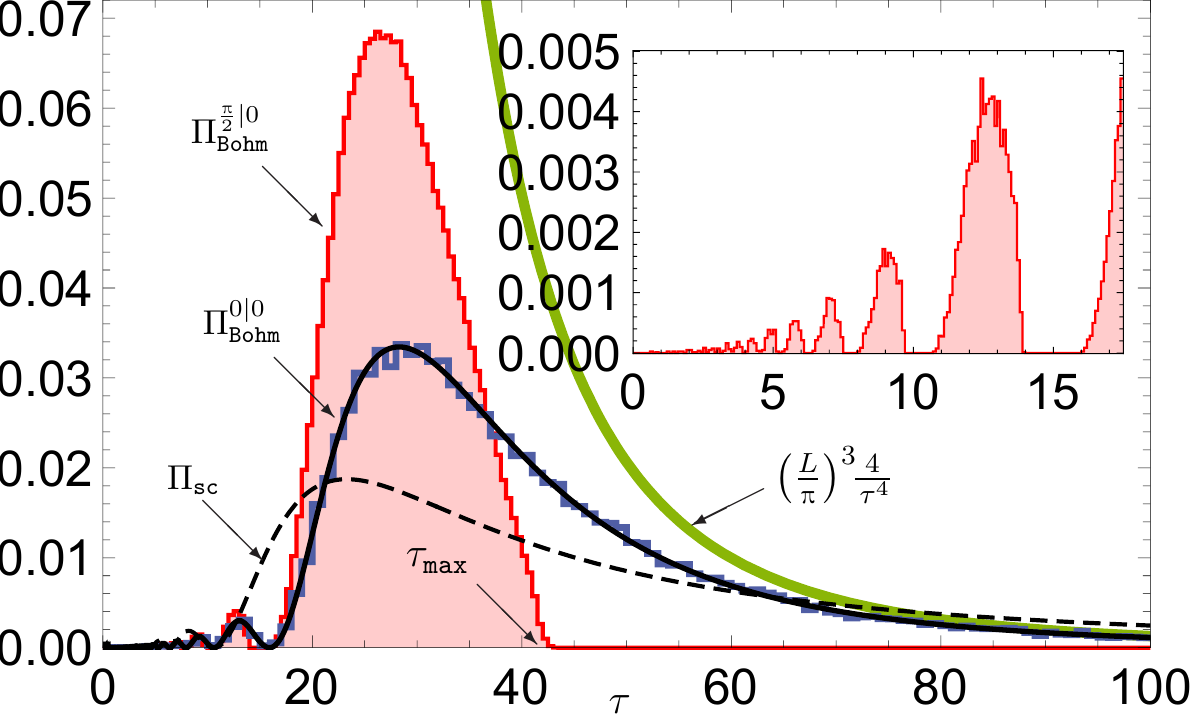}
\caption{Arrival time histograms for spin-up $\big(\Pi^{0|0}_{\texttt{Bohm}}(\tau)\big)$ and up-down $\big(\Pi^{\frac{\pi}{2}|0}_{\texttt{Bohm}}(\tau)\big)$ wave functions, $L=100$ and $\omega=10^3$ graphed along with the semiclassical arrival time distribution $\Pi_{\texttt{sc}}(\tau)$ (dashed line) and the quantum (convective) flux distribution $\Pi_{\texttt{qf}}(\tau)$ (solid line). We see agreement between $\Pi^{0|0}_{\texttt{Bohm}}(\tau)$ and $\Pi_{\texttt{qf}}(\tau)$. For the up-down case, no arrivals are recorded for $\tau>42.9$ ($=\tau_{\texttt{max}}$). Note the disagreement of all distributions with $\Pi_{\texttt{sc}}(\tau)$. Each histogram in this figure has been generated with $10^5$ Bohmian trajectories. The time scale on the horizontal axis is $\approx21.7\,\text{\textmu}\texttt{s}$, assuming $d=50\,\text{\textmu}\texttt{m}$. Inset: Magnified view of the self-similar smaller lobes of the up-down histogram, separated by distinct no-arrival windows.}\label{w1000}
\end{figure}

\par For wave functions corresponding to $0<\alpha\leq\frac{\pi}{2}$ (cf. \eqref{propB}), the first arrival time distribution is \emph{not} given by the integrated flux \eqref{qufl}. This is because the Bohmian trajectories in these cases cross $\partial\Sigma$ more than once, hence the aforementioned current positivity condition is not met. As $\alpha$ approaches $\frac{\pi}{2}$, the tail of $\Pi_{\texttt{Bohm}}^{\alpha|0}(\tau)$ thins gradually, pinching off completely at a characteristic maximum arrival time $\tau_{\texttt{max}}$ for $\alpha=\frac{\pi}{2}$ (i.e. \emph{all} Bohmian trajectories with wave function $\Psi_\ud$ strike the detector surface $z=L$ before $t=\tau_\texttt{max}$). In Fig. \ref{w1000} $\tau_{\texttt{max}}\approx42.9$, which corresponds to $\approx1\,\texttt{ms}$. This behavior results in a sharp drop in the mean first arrival time $\expval{\tau}$ in the vicinity of $\alpha=\frac{\pi}{2}$, as shown in Fig. \ref{mean} below.

\begin{figure}[!ht]
\centering
\includegraphics[scale=1]{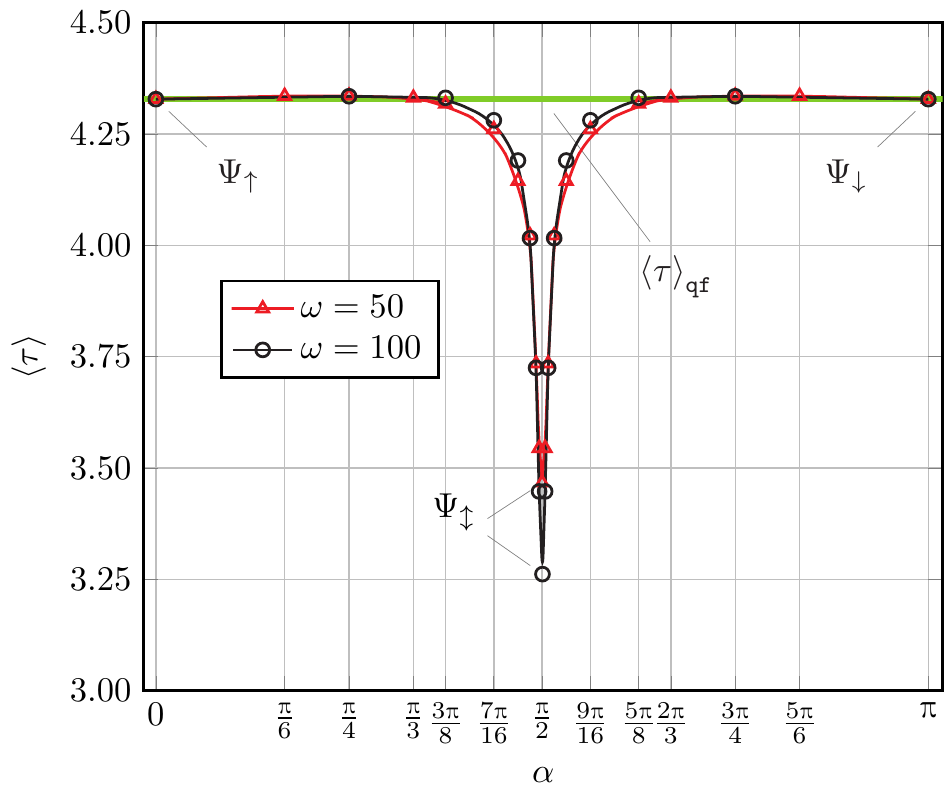}
\caption{Mean first arrival time $\expval{\tau}$ vs. spin orientation angle $\alpha$ for $L=10$ and $\beta=0$. The symmetry of the curves about $\alpha=\frac{\pi}{2}$ is a consequence of property \eqref{propB}.}\label{mean}
\end{figure}

\par
\begin{figure}[!ht]
\hspace{0.5cm}
\begin{subfigure}{.5\textwidth}
  \centering
  \includegraphics[width=6cm]{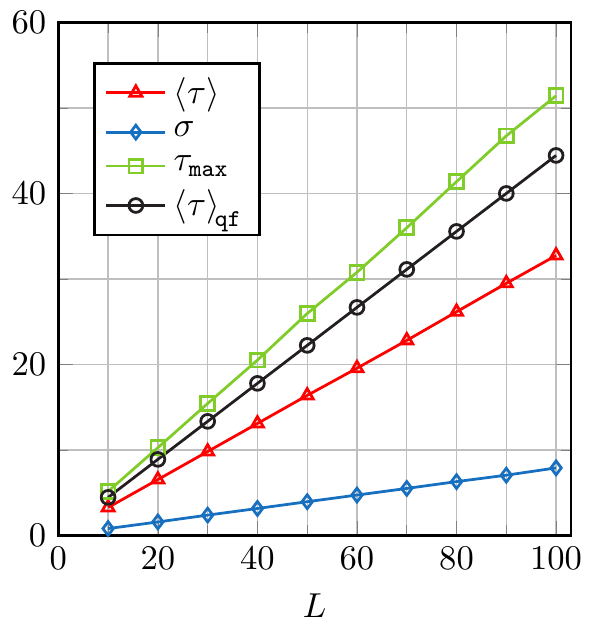}
  \caption{$\omega=100$}
\end{subfigure}%
\hspace{-1.5cm}
\begin{subfigure}{.5\textwidth}
  \centering
  \includegraphics[width=6.75cm]{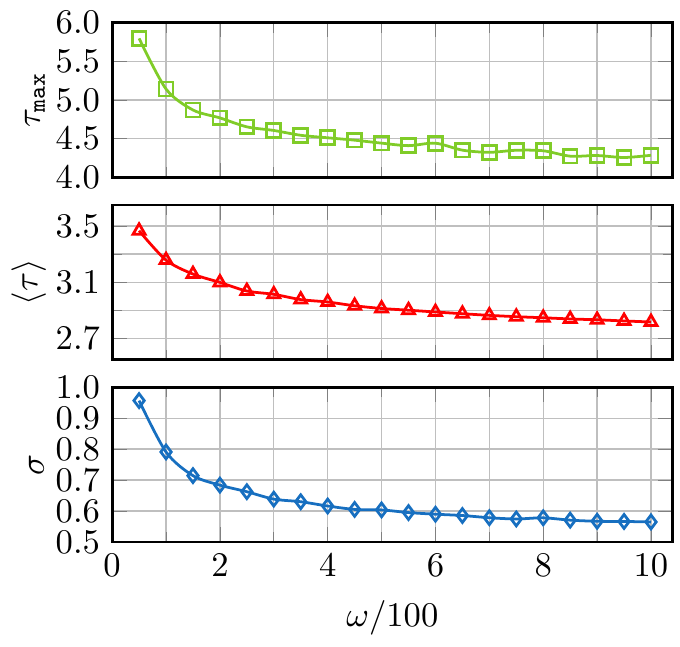}
  \caption{$L=10$}
\end{subfigure}
\caption{(a) Graphs of mean first arrival time $\expval{\tau}$, standard deviation $\sigma$ and maximum arrival time $\tau_{\texttt{max}}$ for the up-down wave function vs. $L$, keeping $\omega$ fixed. The mean arrival time of $\Pi_{\texttt{qf}}(\tau)$ is also shown here. (b) Graphs of mean, standard deviation and maximum arrival time vs. $\omega$, keeping $L$ fixed.}
\label{PihalfStats}
\end{figure}

 Unlike the spin-up (down) case, the up-down arrival time statistics are influenced by the trapping frequency $\omega$. Keeping $L$ fixed, we find that the maximum arrival time $\tau_{\texttt{max}}$, mean first arrival time $\expval{\tau}$, and the standard deviation $\sigma$ corresponding to $\Pi_{\texttt{Bohm}}^{\frac{\pi}{2}|0}(\tau)$ decrease with increasing $\omega$, each approaching a \emph{constant} for $\omega\gg1$. Conversely, when these quantities are graphed as functions of $L$ with $\omega$ fixed, we see a clear linear growth in each (see Fig. \ref{PihalfStats}). 
Remarkably, these effects persist even for large $L$, provided $\omega$ is also made suitably large (a welcome feature for Penning traps). Increasing $L$ causes the arrival time distributions to shift to larger values of $\tau$, thus the smaller lobes can be easily seen, especially in experiments incapable of resolving very small arrival times.

\par Since the smaller (self-similar) lobes become progressively smaller, resolving the $n^{\text{th}}$ lobe (denoting the main lobe by $n=1$) is limited by the least time count ($\delta t$) of the measuring apparatus. Roughly, $\delta t<\text{width of $n^{\text{th}}$ lobe}$ should suffice to resolve the first $n$ lobes. This translates into
\begin{equation}\label{leastcount}
    \delta t<\left(\frac{md}{\pi\hbar}\right)\frac{L}{n^2}.
\end{equation}
For $d=50\,\text{\textmu}\texttt{m}$ and $L=5\,\texttt{mm}$ (Fig. \ref{w1000}), a modest $\delta t\approx10\,\text{\textmu}\texttt{s}$ will successfully resolve $8$ lobes (main + $7$ smaller lobes), while $\delta t\approx0.1\,\text{\textmu}\texttt{s}$ will resolve as many as $83$ lobes (main + $82$ smaller lobes). However, we must also understand that only a few data points (about $\left(\frac{2}{\pi^2}\right)\frac{N}{n^4}$ in $N$ experiments) contribute to the $n^{\text{th}}$ lobe, especially when $n\gg1$. This number, being independent of any tunable parameters like $L$, $\omega$, etc., sets an intrinsic limit on the experimenter's ability to resolve the distant lobes.

\par Finally, we come to the semiclassical arrival time distribution (dashed line in Fig. \ref{w1000}),
\begin{equation}\label{semi}
    \Pi_{\texttt{sc}}(\tau)=\int_{\partial\Sigma}\!\frac{\bb{r}\!\cdot\!\mathrm{d}\bb{\text{s}}}{\tau^4}\left|\tilde{\psi}_0\!\left(\frac{\bb{r}}{\tau}\right)\right|^2,
\end{equation}
routinely used in the interpretation of time-of-flight experiments. Here, $\tilde{\psi}_0$ denotes the Fourier transform of the initial wave function $\psi_0$. Although \eqref{semi} is based on the tacit assumption that the particle \textit{moves classically} between preparation and measurement stages (see \S~5.3.1 of \cite{Vona}), it can also be motivated from the scattering formalism \cite[pg 971]{DDGZ}, provided the detector surface ($\partial\Sigma$) is placed far away from the support of the initial wave function $\psi_0$ (far field regime). In typical cold atom experiments these conditions are met, hence the semiclassical formula \eqref{semi} is empirically adequate. Therefore, soliciting deviations from \eqref{semi}, theorists have recommended ``moving the detectors closer to the region of coherent wave packet production, or closer to the interaction region'' \cite[pg 419]{MUGA1, Delgado} (i.e. $L\approx d$). However, such a relocation may disturb the wave function of the particle in an undesirable way.

For a meaningful comparison with our results, Eq. \eqref{semi} (which is only applicable for free propagation) must be generalized to account for the presence of the waveguide. A careful calculation \cite{prep} yields
\begin{equation}\label{corrected}
    \Pi_{\texttt{sc}}(\tau)=\frac{8\pi L}{\tau^2}\frac{\cos^2(L/2\tau)}{((L/\tau)^2-\pi^2)^2},
\end{equation}
which falls off as $(\frac{8}{\pi^3})\frac{L}{\tau^2}$ (compare this with Eq. \eqref{flux} and its $\tau^{-4}$ fall-off). In Fig. \ref{w1000}, we see that the semiclassical formula \eqref{corrected} for $L=100$ remains \textit{distinctly different} from the Bohmian arrival time distributions, notwithstanding the largeness of $L$. 

\par Bohmian mechanics has been fruitfully applied in various physical disciplines with reference to technical advances \cite{Marian,Wyatt,Oriols,Artes}, and to a better understanding of quantum phenomena, even in quantum gravity \cite{ward}. We have proposed here a simple experiment that can possibly be performed with a detection mechanism as discussed in \cite{Mugaphoton,Allcock3,NICE}, and we have provided the Bohmian arrival times as a benchmark: Our results demonstrate that the distribution of first arrival times of a spin-1/2 particle bears clear signatures of Bohm's guidance law. The deviations of $\Pi_{\texttt{Bohm}}^{\Psi_0}(\tau)$ from the quantum flux distribution \eqref{qufl}, so strikingly in evidence, are particularly noteworthy. 


\begin{thebibliography}{10}
\urlstyle{rm}
\expandafter\ifx\csname url\endcsname\relax
  \def\url#1{\texttt{#1}}\fi
\expandafter\ifx\csname urlprefix\endcsname\relax\def\urlprefix{URL }\fi
\expandafter\ifx\csname doiprefix\endcsname\relax\def\doiprefix{DOI: }\fi
\providecommand{\bibinfo}[2]{#2}
\providecommand{\eprint}[2][]{\url{#2}}

\bibitem{MUGA}
\bibinfo{editor}{Muga J.~G.}, \bibinfo{editor}{Mayato R.~S.} \&
  \bibinfo{editor}{Egusquiza {\'{I}}. L.} (eds.) \emph{\bibinfo{title}{Time in
  Quantum Mechanics}}, vol.~\bibinfo{volume}{1} of \emph{\bibinfo{series}{Lect.
  Notes Phys. 734}} (\bibinfo{publisher}{Springer}, \bibinfo{address}{Berlin
  Heidelberg}, \bibinfo{year}{2008}), \bibinfo{edition}{second} edn.

\bibitem{MUGA1}
\bibinfo{author}{Muga J.~G.} \& \bibinfo{author}{Leavens C.~R.}
\newblock \bibinfo{journal}{\bibinfo{title}{Arrival time in quantum mechanics}}.
\newblock {\emph{\JournalTitle{Phys. Rep.}}} \textbf{\bibinfo{volume}{338}},
  \bibinfo{pages}{353--438}, \doiprefix\url{10.1016/S0370-1573(00)00047-8}
  (\bibinfo{year}{2000}).

\bibitem{Allcock1}
\bibinfo{author}{Allcock G.~R.}
\newblock \bibinfo{journal}{\bibinfo{title}{The time of arrival in quantum mechanics I--Formal considerations}}.
\newblock {\emph{\JournalTitle{Ann. Phys.}}} \textbf{\bibinfo{volume}{53}},
  \bibinfo{pages}{253--285}, \doiprefix\url{10.1016/0003-4916(69)90251-6}
  (\bibinfo{year}{1969}).

\bibitem{AhBohm}
\bibinfo{author}{Aharonov Y.} \& \bibinfo{author}{Bohm D.}
\newblock \bibinfo{journal}{\bibinfo{title}{Time in the quantum theory and the uncertainty relation for time and energy}}.
\newblock {\emph{\JournalTitle{Phys. Rev.}}} \textbf{\bibinfo{volume}{122}},
  \bibinfo{pages}{1649--1658}, \doiprefix\url{10.1103/PhysRev.122.1649}
  (\bibinfo{year}{1961}).

\bibitem{Vona}
\bibinfo{editor}{Blanchard P.} \& \bibinfo{editor}{Fr{\"{o}}hlich J.} (eds.)
 \emph{\bibinfo{title}{The Message of Quantum Science: Attempts Towards a Synthesis}}.
\newblock Lect. Notes Phys. 899 (\bibinfo{publisher}{Springer},
  \bibinfo{address}{Berlin Heidelberg}, \bibinfo{year}{2015}).

\bibitem{DDGZ}
\bibinfo{author}{Daumer M.}, \bibinfo{author}{D{\"{u}}rr D.},
  \bibinfo{author}{Goldstein S.} \& \bibinfo{author}{Zangh{\`{I}} N.}
\newblock \bibinfo{journal}{\bibinfo{title}{On the quantum probability flux through surfaces}}.
\newblock {\emph{\JournalTitle{J. Stat. Phys.}}} \textbf{\bibinfo{volume}{88}},
  \bibinfo{pages}{967--–977},
  \doiprefix\url{10.1023/B:JOSS.0000015181.86864.fb} (\bibinfo{year}{1997}).

\bibitem{Vona1}
\bibinfo{author}{Vona N.}, \bibinfo{author}{Hinrichs G.} \&
  \bibinfo{author}{D\"urr D.}
\newblock \bibinfo{journal}{\bibinfo{title}{What does one measure when one measures the arrival time of a quantum particle?}}
\newblock {\emph{\JournalTitle{Phys. Rev. Lett.}}}
  \textbf{\bibinfo{volume}{111}}, \bibinfo{pages}{220404},
  \doiprefix\url{10.1103/PhysRevLett.111.220404} (\bibinfo{year}{2013}).

\bibitem{Rodi}
\bibinfo{author}{Tumulka R.}
\newblock \bibinfo{journal}{\bibinfo{title}{Distribution of the time at which
  an ideal detector clicks}}.
\newblock {\emph{\JournalTitle{ArXiv e-prints}}}  (\bibinfo{year}{2016}).
\newblock \eprint{1601.03715}.

\bibitem{DGZOperators}
\bibinfo{author}{D{\"{u}}rr D.}, \bibinfo{author}{Goldstein S.} \&
  \bibinfo{author}{Zangh{\`{I}} N.}
\newblock \bibinfo{journal}{\bibinfo{title}{Quantum equilibrium and the role of operators as observables in quantum theory}}.
\newblock {\emph{\JournalTitle{J. Stat. Phys.}}}
\textbf{\bibinfo{volume}{116}}, \bibinfo{pages}{959--–1055},
\doiprefix\url{10.1023/B:JOSS.0000} (\bibinfo{year}{2004}).

\bibitem{Kocsis}
\bibinfo{author}{Kocsis S.} \emph{et~al.}
\newblock \bibinfo{journal}{\bibinfo{title}{Observing the average trajectories
  of single photons in a two-slit interferometer}}.
\newblock {\emph{\JournalTitle{Science}}} \textbf{\bibinfo{volume}{332}},
  \bibinfo{pages}{1170--1173}, \doiprefix\url{10.1126/science.1202218}
  (\bibinfo{year}{2011}).

\bibitem{Yearsley}
\bibinfo{author}{Yearsley J.~M.}
\newblock \bibinfo{journal}{\bibinfo{title}{A review of the decoherent
  histories approach to the arrival time problem in quantum theory}}.
\newblock {\emph{\JournalTitle{J. Phys. Conf. Ser}}}
  \textbf{\bibinfo{volume}{306}}, \bibinfo{pages}{012056},
  \doiprefix\url{10.1088/1742-6596/306/1/012056} (\bibinfo{year}{2011}).

\bibitem{Allcock2}
\bibinfo{author}{Allcock G.~R.}
\newblock \bibinfo{journal}{\bibinfo{title}{The time of arrival in quantum
  mechanics II--The individual measurement}}.
\newblock {\emph{\JournalTitle{Ann. Phys.}}} \textbf{\bibinfo{volume}{53}},
  \bibinfo{pages}{286--310}, \doiprefix\url{10.1016/0003-4916(69)90252-8}
  (\bibinfo{year}{1969}).

\bibitem{Allcock3}
\bibinfo{author}{Allcock G.~R.}
\newblock \bibinfo{journal}{\bibinfo{title}{The time of arrival in quantum
  mechanics III--The measurement ensemble}}.
\newblock {\emph{\JournalTitle{Ann. Phys.}}} \textbf{\bibinfo{volume}{53}},
  \bibinfo{pages}{311--348}, \doiprefix\url{10.1016/0003-4916(69)90253-X}
  (\bibinfo{year}{1969}).

\bibitem{Keller}
\bibinfo{author}{Landsman A.~S.} \emph{et~al.}
\newblock \bibinfo{journal}{\bibinfo{title}{Ultrafast resolution of tunneling delay time}}.
\newblock {\emph{\JournalTitle{Optica}}} \textbf{\bibinfo{volume}{1}},
  \bibinfo{pages}{343--349}, \doiprefix\url{10.1364/OPTICA.1.000343}
  (\bibinfo{year}{2014}).

\bibitem{Lundsmann}
\bibinfo{author}{Zimmermann T.} \emph{et~al.}
\newblock \bibinfo{journal}{\bibinfo{title}{Tunneling time and weak measurement in strong field ionization}}.
\newblock {\emph{\JournalTitle{Phys. Rev. Lett.}}}
  \textbf{\bibinfo{volume}{116}}, \bibinfo{pages}{233603},
  \doiprefix\url{10.1103/PhysRevLett.116.233603} (\bibinfo{year}{2016}).

\bibitem{BohmHiley}
\bibinfo{author}{Bohm D.} \& \bibinfo{author}{Hiley B.~J.}
\newblock \emph{\bibinfo{title}{The Undivided Universe: An Ontological
  Interpretation of Quantum Theory}} (\bibinfo{publisher}{Routledge},
  \bibinfo{address}{London and New York}, \bibinfo{year}{1993}).

\bibitem{Leavens}
\bibinfo{author}{Leavens C.~R.}
\newblock \bibinfo{journal}{\bibinfo{title}{Time of arrival in quantum and Bohmian mechanics}}.
\newblock {\emph{\JournalTitle{Phys. Rev. A}}} \textbf{\bibinfo{volume}{58}},
  \bibinfo{pages}{840--847}, \doiprefix\url{10.1103/PhysRevA.58.840}
  (\bibinfo{year}{1998}).

\bibitem{J1}
\bibinfo{author}{Shikakhwa M.~S.}, \bibinfo{author}{Turgut S.} \&
  \bibinfo{author}{Pak N.~K.}
\newblock \bibinfo{journal}{\bibinfo{title}{Derivation of the magnetization current from the non-relativistic Pauli equation}}.
\newblock {\emph{\JournalTitle{Am. J. Phys.}}} \textbf{\bibinfo{volume}{79}},
  \bibinfo{pages}{1177--1179}, \doiprefix\url{10.1119/1.3630931}
  (\bibinfo{year}{2011}).

\bibitem{J2}
\bibinfo{author}{Hodge W.~B.}, \bibinfo{author}{Migirditch S.~V.} \&
  \bibinfo{author}{Kerr W.~C.}
\newblock \bibinfo{journal}{\bibinfo{title}{Electron spin and probability
  current density in quantum mechanics}}.
\newblock {\emph{\JournalTitle{Am. J. Phys.}}} \textbf{\bibinfo{volume}{82}},
  \bibinfo{pages}{681--690}, \doiprefix\url{10.1119/1.4868094}
  (\bibinfo{year}{2014}).

\bibitem{Holland2003}
\bibinfo{author}{Holland P.~R.}
\newblock \bibinfo{journal}{\bibinfo{title}{Uniqueness of conserved currents in
  quantum mechanics}}.
\newblock {\emph{\JournalTitle{Ann. Phys. (Leipzig)}}}
  \textbf{\bibinfo{volume}{12}}, \bibinfo{pages}{446--462},
  \doiprefix\url{10.1002/andp.200310022} (\bibinfo{year}{2003}).

\bibitem{Holland}
\bibinfo{author}{Holland P.~R.}
\newblock \bibinfo{journal}{\bibinfo{title}{Uniqueness of paths in quantum
  mechanics}}.
\newblock {\emph{\JournalTitle{Phy. Rev. A}}} \textbf{\bibinfo{volume}{60}},
  \bibinfo{pages}{4326--4330}, \doiprefix\url{10.1103/PhysRevA.60.4326}
  (\bibinfo{year}{1999}).

\bibitem{HollandPhilippidis}
\bibinfo{author}{Holland P.~R.} \& \bibinfo{author}{Philippidis C.}
\newblock \bibinfo{journal}{\bibinfo{title}{Implications of Lorentz covariance
  for the guidance equation in two-slit quantum interference}}.
\newblock {\emph{\JournalTitle{Phy. Rev. A}}} \textbf{\bibinfo{volume}{67}},
  \bibinfo{pages}{062105}, \doiprefix\url{10.1103/PhysRevA.67.062105}
  (\bibinfo{year}{2003}).

\bibitem{HDK1}
\bibinfo{author}{Dewdney C.}, \bibinfo{author}{Holland P.~R.} \&
  \bibinfo{author}{Kyprianidis C.}
\newblock \bibinfo{journal}{\bibinfo{title}{What happens in a spin
  measurement?}}
\newblock {\emph{\JournalTitle{Phys. Lett. A}}} \textbf{\bibinfo{volume}{119}},
  \bibinfo{pages}{259--267}, \doiprefix\url{10.1016/0375-9601(86)90144-1}
  (\bibinfo{year}{1986}).

\bibitem{HDK2}
\bibinfo{author}{Dewdney C.}, \bibinfo{author}{Holland P.~R.},
  \bibinfo{author}{Kyprianidis C.} \& \bibinfo{author}{Vigier J.~P.}
\newblock \bibinfo{journal}{\bibinfo{title}{Spin and non-locality in quantum mechanics}}.
\newblock {\emph{\JournalTitle{Nature}}} \textbf{\bibinfo{volume}{336}},
  \bibinfo{pages}{536--544}, \doiprefix\url{10.1038/336536a0}
  (\bibinfo{year}{1988}).

\bibitem{DGZBorn}
\bibinfo{author}{D{\"{u}}rr D.}, \bibinfo{author}{Goldstein S.} \&
  \bibinfo{author}{Zangh{\`{I}} N.}
\newblock \bibinfo{journal}{\bibinfo{title}{Quantum equilibrium and the origin
  of absolute uncertainty}}.
\newblock {\emph{\JournalTitle{J. Stat. Phys.}}} \textbf{\bibinfo{volume}{67}},
  \bibinfo{pages}{843–--907}, \doiprefix\url{10.1007/BF01049004}
  (\bibinfo{year}{1992}).

\bibitem{Dehmelt}
\bibinfo{author}{Wineland D.}, \bibinfo{author}{Ekstrom P.} \&
  \bibinfo{author}{Dehmelt H.}
\newblock \bibinfo{journal}{\bibinfo{title}{Monoelectron oscillator}}.
\newblock {\emph{\JournalTitle{Phys. Rev. Lett.}}}
  \textbf{\bibinfo{volume}{31}}, \bibinfo{pages}{1279--1282},
  \doiprefix\url{10.1103/PhysRevLett.31.1279} (\bibinfo{year}{1973}).

\bibitem{Joseph}
\bibinfo{author}{Joseph T.} \& \bibinfo{author}{Gabrielse G.}
\newblock \bibinfo{journal}{\bibinfo{title}{One electron in an orthogonalized
  cylindrical Penning trap}}.
\newblock {\emph{\JournalTitle{Appl. Phys. Lett.}}}
  \textbf{\bibinfo{volume}{55}}, \bibinfo{pages}{2144--2146},
  \doiprefix\url{10.1063/1.102084} (\bibinfo{year}{1989}).

\bibitem{Ulmer}
\bibinfo{author}{Ulmer S.} \emph{et~al.}
\newblock \bibinfo{journal}{\bibinfo{title}{Observation of spin flips with a
  single trapped proton}}.
\newblock {\emph{\JournalTitle{Phys. Rev. Lett.}}}
  \textbf{\bibinfo{volume}{106}}, \bibinfo{pages}{253001},
  \doiprefix\url{10.1103/PhysRevLett.106.253001} (\bibinfo{year}{2011}).

\bibitem{Dehmelt1}
\bibinfo{author}{Dehmelt H.~G.}
\newblock \bibinfo{journal}{\bibinfo{title}{Nobel lecture: Experiments with an
  isolated subatomic particle at rest}}.
\newblock {\emph{\JournalTitle{Rev. Mod. Phys.}}}
  \textbf{\bibinfo{volume}{62}}, \bibinfo{pages}{525--530},
  \doiprefix\url{10.1103/RevModPhys.62.525} (\bibinfo{year}{1990}).

\bibitem{DMD}
\bibinfo{author}{Das S.}, \bibinfo{author}{N{\"o}th M.} \&
\bibinfo{author}{D{\"u}rr D.}
\newblock \bibinfo{journal}{\bibinfo{title}{Exotic Bohmian arrival times of spin-1/2 particles (in preparation)}}.

\bibitem{Moshinsky}
\bibinfo{author}{Moshinsky M.}
\newblock \bibinfo{journal}{\bibinfo{title}{Diffraction in time}}.
\newblock {\emph{\JournalTitle{Phys. Rev.}}} \textbf{\bibinfo{volume}{88}},
  \bibinfo{pages}{625--631}, \doiprefix\url{10.1103/PhysRev.88.625}
  (\bibinfo{year}{1952}).

\bibitem{prep}
\bibinfo{author}{Das S.}
\newblock \emph{\bibinfo{title}{Arrival Time Distributions of Spin-1/2 Particles}}.
\newblock Master's thesis, \bibinfo{school}{LMU Munich {\&} TU Munich}
  (\bibinfo{year}{2017}).
\newblock
\bibinfo{note}{\url{http://www.mathematik.uni-muenchen.de/~bohmmech/theses/Das_Siddhant_MA.pdf} and S. Das {\&} D. D{\"{u}}rr in preparation}.

\bibitem{Delgado}
\bibinfo{author}{Delgado V.}
\newblock \bibinfo{journal}{\bibinfo{title}{Quantum probability distribution of arrival times and probability current density}}.
\newblock {\emph{\JournalTitle{Phys. Rev. A}}} \textbf{\bibinfo{volume}{59}},
\bibinfo{pages}{1010--1020}, \doiprefix\url{10.1103/PhysRevA.59.1010} (\bibinfo{year}{1999}).

\bibitem{Marian}
\bibinfo{author}{Marian D.}, \bibinfo{author}{Zangh\`{\i} N.} \&
  \bibinfo{author}{Oriols X.}
\newblock \bibinfo{journal}{\bibinfo{title}{Weak values from displacement
  currents in multiterminal electron devices}}.
\newblock {\emph{\JournalTitle{Phys. Rev. Lett.}}}
  \textbf{\bibinfo{volume}{116}}, \bibinfo{pages}{110404},
  \doiprefix\url{10.1103/PhysRevLett.116.110404} (\bibinfo{year}{2016}).

\bibitem{Wyatt}
\bibinfo{author}{Wyatt R.~E.}
\newblock \emph{\bibinfo{title}{Quantum Dynamics with Trajectories: Introduction to Quantum Hydrodynamics}} (\bibinfo{publisher}{Springer},
  \bibinfo{address}{New York}, \bibinfo{year}{2005}).

\bibitem{Oriols}
\bibinfo{editor}{Oriols X.} \& \bibinfo{editor}{Mompart J.} (eds.)
  \emph{\bibinfo{title}{Applied Bohmian Mechanics: From Nanoscale Systems to
  Cosmology}} (\bibinfo{publisher}{Pan Stanford Publishing Pvt. Ltd.},
  \bibinfo{address}{Singapore}, \bibinfo{year}{2012}).

\bibitem{Artes}
\bibinfo{editor}{Sanz A.~S.} \& \bibinfo{editor}{Miret-Art{\'{e}}s S.} (eds.)
  \emph{\bibinfo{title}{A Trajectory Description of Quantum Processes}},
  vol.~\bibinfo{volume}{2} of \emph{\bibinfo{series}{Lect. Notes Phys. 831}}
  (\bibinfo{publisher}{Springer}, \bibinfo{address}{Berlin Heidelberg},
  \bibinfo{year}{2014}), \bibinfo{edition}{second} edn.

\bibitem{ward}
\bibinfo{author}{Struyve W.}
\newblock \bibinfo{journal}{\bibinfo{title}{Loop quantum cosmology \&
  singularities}}.
\newblock {\emph{\JournalTitle{Sci. Rep.}}} \textbf{\bibinfo{volume}{7}},
  \doiprefix\url{10.1038/s41598-017-06616-y} (\bibinfo{year}{2017}).

\bibitem{Mugaphoton}
\bibinfo{author}{Damborenea J.~A.}, \bibinfo{author}{Egusquiza {\'{I}}.~L.},
  \bibinfo{author}{Hegerfeldt G.~C.} \& \bibinfo{author}{Muga J.~G.}
\newblock \bibinfo{journal}{\bibinfo{title}{Measurement-based approach to quantum arrival times}}.
\newblock {\emph{\JournalTitle{Physical Review. A}}}
  \textbf{\bibinfo{volume}{66}}, \doiprefix\url{10.1103/PhysRevA.66.052104}
  (\bibinfo{year}{2002}).

\bibitem{NICE}
\bibinfo{author}{Uehara Y.} \emph{et~al.}
\newblock \bibinfo{journal}{\bibinfo{title}{High resolution time-of-flight electron spectrometer}}.
\newblock {\emph{\JournalTitle{Jpn. J. Appl. Phys}}}
  \textbf{\bibinfo{volume}{29}}, \bibinfo{pages}{2858--2863},
  \doiprefix\url{10.1143/JJAP.29.2858} (\bibinfo{year}{1990}).

\end{thebibliography}

\section*{Acknowledgements}

The authors would like to thank J. M. Wilkes for critically reviewing the manuscript. He and Markus N\"{o}th verified our analytical calculations meticulously, which led to the correction of a mistake in the early stages of the work. The assistance of Grzesio Gradziuk and Leopold Kellers with the numerical simulations was essential, and greatly appreciated. Thanks are also due to Serj Aristarkhov, Lukas Nickel, Ward Struyve, Nikolai Leopold, Roderich Tumulka and Nicola Vona for inspiration and enriching discussions on the subject of this letter.

\section*{Author contributions statement}
S.D. conceived the experiment and conducted the numerical computations. The paper was written equally by S.D. and D.D.

\section*{Additional information}

\textbf{Competing interests}: The authors declare no competing interests. \\[-2pt]

\noindent \textbf{Data availability statement}: The datasets generated and analyzed during the current study are available from the corresponding author on reasonable request.

\end{document}